\documentstyle[psfig]{mn}
\begin{document}

\author[J. Falc\'on-Barroso, R.F. Peletier, M. Balcells]
{Jes\'us Falc\'on-Barroso$^1$, Reynier F. Peletier$^{1,2}$, Marc Balcells$^3$
\\$^1$School of Physics \& Astronomy. University of Nottingham. Nottingham. NG7 2RD. United Kingdom
\\$^2$CRAL, Observatoire de Lyon, F-69561 St-Genis Laval cedex, France
\\$^3$Instituto de Astrof\'\i sica de Canarias, E-38200 La Laguna, Tenerife, Spain}
\date{Accepted ... Received ...; in original form}
\title{Bulges on the Fundamental Plane of Early-Type Galaxies}
\maketitle
\begin{abstract}
In an ongoing effort to study the formation and evolution of galactic bulges we
have investigated the position of 19 bulges of type S0-Sbc on the fundamental 
plane (FP). We find that bulges, both in $B$ and $K$ band, lie close to but
slightly below the FP defined by ellipticals and S0s, i.e. are slightly brighter. There are
hints that bulges of later morphological type are situated below the other
bulges in our sample. The FP results are consistent with the picture,  obtained from our
recent analysis of HST-colours, that bulges are old,  except for the Sbc
galaxies. The fact that bulges lie so close to the FP of  ellipticals and S0s
implies that their formation epoch must have been  similar, at most about 2.5 Gyr
earlier, than cluster Es and S0s, and and that the surrounding disk does not significantly 
affect their structure. 
\end{abstract}


\section{Introduction}
\label{Sec:Introduction}
It has been known for almost 15 years that early-type galaxies obey a tight
relation linking their structure with their kinematics. Originally discovered
by Djorgovski \& Davis \shortcite{dd87} and Dressler et al. \shortcite{d87},
the Fundamental Plane (FP) establishes that the effective radius ($r_{e}$), the
mean effective surface brightness ($\Sigma_{e}$) and the central velocity
dispersion ($\sigma_{0}$) are related by

\begin{center}
$r_{e} = C \cdot \sigma_{0}^{A} \cdot \Sigma_{e}^{B}$
\end{center}

The FP is one of the main scaling relations for early-type galaxies (e.g. Pahre,
Djorgovski \& de Carvalho 1998, hereafter P98). The fact that, compared to
other relations, its scatter is so small, has made it a useful distance indicator
(J\o rgensen, Franx \& Kj\ae rgaard 1996, hereafter J96).
Historically, the FP is a refinement of the Faber-Jackson \shortcite{fj76}, 
relation between galaxy luminosity and $\sigma_0$, which includes the dependency
of luminosity and $\sigma_0$ on surface brightness (Kormendy \shortcite{k85}).

Much work has been devoted to determining the exponents $A$ and $B$. The virial
theorem predicts $A=2$ and $B=-1$. However, observations show lower values of
$A$ and higher values of $B$ (Busarello et al. 1997). The value $B$ has been
well determined to be $\sim -0.8$ (P98, J96). $A$, usually referred to as the
'tilt' of the FP (P98), has values that vary significantly depending on the
band employed. Several mechanisms are likely to produce the difference between
observations and the Virial Theorem prediction: age and metallicity, shape of
the light profiles \cite{c93,g97,ps94,ps97} variations in the initial mass function
(IMF) or even the dark matter distribution in galaxies \cite{g93,c96}.

Beyond its value as a distance
indicator,  the FP contains useful clues on the physics of galactic systems.  
The position of galactic bulges on the FP of
elliptical galaxies provides information about the differences in
stellar populations, internal structure and kinematics of both types of
galaxies. In a series of papers, Prugniel \& Simien (1994, 1996, 1997) have
gone
into the details of using the FP for this purpose. Assuming that the FP is
a representation of the Virial Theorem $\sigma_0^2 ~ \propto ~ {M \over r_e}$,
and that the mass-to-light ratio scales as ${M \over L} \propto L^\beta$,
these authors obtain the relation between observables $\sigma_0^2 ~ \propto ~
{1 \over
r_e} L^{1+\beta}$. In their picture, this relation is tight for old elliptical
galaxies, since for them $\sigma_0$ is a good mass-indicator, and $M\over L$ is
a smooth function of L, probably metallicity-driven. Deviations to the relation
can be due to the presence of young stellar populations (causing a change in 
$M\over L$, see also Schweizer \& Seitzer 1992), rotational support (since in
this case $\sigma_0$ gives an underestimate of the gravitational  energy), or
differences in
structure, for example indicated by the shape of the surface brightness
profile.

One of the first papers addressing the position of bulges on the FP is the 
study  of Bender, Burstein \& Faber (1992, 
hereafter B92).  They present $B$-band FP properties for a large sample
of dynamically hot galaxies, which they divide into several groups, referred
to as Es, intermediate Es, compact Es, bulges, bright dwarf ellipticals and
dwarf spheroidals. B92 conclude that bulges lie preferentially below the FP of 
ellipticals, suggesting that, on average, bulges are younger than ellipticals.
Most (17 out of 19) of the bulges in B92 are in lenticular galaxies. In a related
study using a sample of galaxies from the Coma cluster Saglia, Bender \&
Dressler \shortcite{s93} find a small shift in the same direction in the
location of bulges when compared with Es and S0s (0.1 in log($r_{e}$) in B-band).
In contrast with these findings, J96, in their comparison between the FP
of 10 clusters, find  a negligible difference between E and S0 galaxies
($\Delta(E-S0)=0.010 \pm 0.015$ in B-band), a result shared by
Prugniel \& Simien \shortcite{ps96}.

The FP of bulges is more complicated to study than that of elliptical galaxies,
since a determination of structural parameters requires a decomposition of the
light into bulges and disks, as well as to take into account dust reddening and
extinction. These problems only worsen as we move to later types. Perhaps as a
result of this, very little information is available about the FP of bulges  of
types later than S0. This paper intends to fill this gap by presenting a study
of the FP of bulges for types ranging from S0 to Sbc. It 
belongs to  a
series of papers discussing the properties of  galactic bulges. The first paper
\cite{bp94} discusses optical colours and colour gradients of an optically
selected, complete, sample of $\sim$ 50 bulges of type S0-Sbc. The galaxies
were required to have an inclination i $>$50$^{\rm o}$ to minimize the amount
of disk light which projects onto the bulge and ensure that extinction from
dust in the disk does not dominate the colours measured on one side of the
bulge. Of this sample, 30 galaxies for which the colour profiles were smooth
on one side of the bulge (i.e. where the colours were not significantly
affected by dust extinction) were observed in the near-infrared $K$-band at
higher spatial resolution. Andredakis, Peletier \& Balcells (1995, hereafter
APB95) present a  photometric decomposition into bulge and disk, using Kent's
(1986) model-independent method, after which S\'ersic r$^{1/n}$ laws are fitted
to the bulge. They find that  the S\'ersic law generally fits very well, and
that there is a good correlation between $n$ and morphological type, such that
bulges of early-type spiral galaxies follow the r$^{1/4}$ law, while profiles
of late type spirals are more nearly exponential. Optical and near-infrared
colours of bulges are presented in Peletier \& Balcells \shortcite{p96}.
Photometric data for this sample, including colour profiles along major and
minor axis and fits of ellipticity and position angle as a function of radius,
are presented in Peletier \& Balcells (1997, hereafter PB97). Central color
profiles, based on HST/WFPC2 and NICMOS imaging of a subsample of 20 of our
galaxies, are analysed in Peletier et al. \shortcite{p99}. The latter paper
concludes that all bulges, except those of Sbc's, are old with an age spread of
less than 2 Gyr. The present paper makes use of our own spectroscopy of the
sample to address the position of bulges on the FP of elliptical galaxies. A
full report  of the kinematic properties derived from the spectroscopy will be 
given in a forthcoming paper.

When analyzing bulges, several authors have made the assumption that bulges'
surface brightness profiles are either $r^{1/4}$ or exponential \cite{cor96}, 
a dichotomy of shapes that would suggest a
dichotomy of formation processes. Carollo \shortcite{carollo99} also divides
bulges into $r^{1/4}$ and exponential classes in her study of galaxy nuclei.
In our work we define and characterize a bulge on the basis of its signature
in the galaxy's ellipticity profile, rather than on the shape of the surface
brightness profile. We thus avoid having to assign each bulge to a given profile
class, and indeed we find a continuum of profile shapes, which we model with
S\'ersic's law with free index $n$, rather than two separate classes. Because
our sample includes objects ranging from exponential to $r^{1/4}$, our FP study
should uncover any structural and population signatures that depend on the
shape of the surface brightness profile.  
 
Section \S2 summarises the observations and data reduction as well as
procedures to derive the different structural and kinematic parameters. 
In \S3 we show the FP in the different bands and compare the results
with other authors.  The Mg$_{2}$ - $\sigma$ relation is analysed in
\S4.  Possible correlations between observables and the residuals of
the FP are studied in \S5.  Throughout the paper we use a Hubble
Constant of $H_{0}$=50 km\,s$^{-1}$Mpc$^{-1}$.

\section{Observations \& Data Reduction}
\subsection{Observations}
\label{Sec:Observations}
We obtained long-slit spectra of our sample galaxies at the 4.2m William
Herschel Telescope of the Observatorio del Roque de los Muchachos at La Palma
between 11 and 13 July 1997. Nineteen galaxies were observed (see Table~1). 
All galaxies were observed between 3660 - 5560 \AA\ and between 8360 - 9170 \AA\
using the ISIS blue and red arms, respectively. The red arm was equipped with the
Tektronix (1024$\times$1024) TEK2 CCD (0.36" per 24 $\mu$m pixel) and the R600R
grating, giving a spectral resolution (FWHM of arc lines) of 1.74 \AA (60 km/s).  
The blue arm was equipped with the Loral (2048$\times$2048) LOR2 CCD, 
(15 $\mu$m pixels, 0.22 "/pixel) and the R300B grating, giving a 
spectral resolution of 4.1 \AA. We used the 6100 \AA\ dichroic for 
simultaneous red and blue arm exposures. 
Typical exposure times per galaxy were 1500 sec in both
wavelength ranges. The slit width was 1.2", matching the seeing at the time of
the observations. Arc line exposures were taken before and after each target
exposure. Tungsten continuum lamp exposures were taken with the red arm after
each target exposure, for fringe calibration. Twilight sky exposures were taken
every night for flat fielding.  We took spectra of spectral templates of types
A2 to M5, from the Lick list of stars \cite{w94}. 

\begin{table*}
\begin{center}
\caption{The Data Set.}
\label{Tab:DataSet}
{\tabcolsep=5.pt
\begin{tabular}{ccccccccccccc}
\hline
\hline
NGC & V$_{\rm{LG}}$ & Scale & $r_{e,K}$ & $<\mu>_{e,K}$  & $r_{e,B}$ & $<\mu>_{e,B}$ & $\log(\sigma_0/{\rm km s^{-1}})$ &S/N& $n$ & T & $\epsilon_{disc}$ & Mg$_{2,0}$\\
~&km/s& kpc/arcsec & arcsec & mag~arsec$^{-2}$ & arcsec & mag~arsec$^{-2}$ & ~ & pix$^-1$  & ~ & ~ & ~ & mag\\
(1) & (2) & (3) & (4) & (5) & (6) & (7) & (8) & (9) & (10) & (11) & (12) & (13)\\
\hline
5326 & 2576 & 0.25 &  1.87  & 13.88 &  2.40  & 18.14 &  2.216  & 84  &  2.19  &  1 & 0.55 & 0.291\\    
  ~  &  ~   &  ~   &   ~    &   ~   &   ~    &   ~   & (0.016) &  ~  & (0.45) &  ~ &  ~   &  ~   \\    
5389 & 1990 & 0.19 &  3.23  & 14.94 &  3.98  & 19.19 &  2.056  & 62  &  3.07  &  0 & 0.75 & 0.275\\    
  ~  &  ~   &  ~   &   ~    &   ~   &   ~    &   ~   & (0.023) &  ~  & (0.24) &  ~ &  ~   &  ~   \\    
5422 & 1929 & 0.19 &  3.63  & 14.78 &  4.44  & 18.97 &  2.205  & 69  &  3.07  & -2 & 0.80 & 0.307\\    
  ~  &  ~   &  ~   &   ~    &   ~   &   ~    &   ~   & (0.016) &  ~  & (0.20) &  ~ &  ~   &  ~   \\    
5443 & 2060 & 0.20 &  3.85  & 15.69 &  4.94  & 19.80 &  1.881  & 38  &  2.86  &  3 & 0.72 & 0.240\\    
  ~  &  ~   &  ~   &   ~    &   ~   &   ~    &   ~   & (0.046) &  ~  & (0.35) &  ~ &  ~   &  ~   \\    
5475 & 1861 & 0.18 &  2.43  & 14.83 &  3.12  & 18.91 &  1.960  & 57  &  2.52  &  1 & 0.71 & 0.239\\    
  ~  &  ~   &  ~   &   ~    &   ~   &   ~    &   ~   & (0.029) &  ~  & (0.25) &  ~ &  ~   &  ~   \\    
5587 & 2291 & 0.22 &  1.76  & 15.13 &  2.05  & 19.02 &  1.967  & 36  &  1.53  &  0 & 0.70 & 0.259\\    
  ~  &  ~   &  ~   &   ~    &   ~   &   ~    &   ~   & (0.038) &  ~  & (0.21) &  ~ &  ~   &  ~   \\  
5689 & 2290 & 0.22 &  6.86  & 15.39 &  9.77  & 20.04 &  2.155  & 59  &  5.90  &  0 & 0.75 & 0.275\\  
  ~  &  ~   &  ~   &   ~    &   ~   &   ~    &   ~   & (0.018) &  ~  & (0.62) &  ~ &  ~   &  ~   \\  
5707 & 2354 & 0.23 &  3.10  & 14.71 &  3.49  & 18.50 &  2.149  & 52  &  1.30  &  2 & 0.75 & 0.257\\  
  ~  &  ~   &  ~   &   ~    &   ~   &   ~    &   ~   & (0.019) &  ~  & (0.27) &  ~ &  ~   &  ~   \\  
5719 & 1684 & 0.16 &  2.77  & 14.69 &  4.47  & 19.98 &  2.034  & 52  &  2.26  &  2 & 0.68 & 0.212\\  
  ~  &  ~   &  ~   &   ~    &   ~   &   ~    &   ~   & (0.024) &  ~  & (0.10) &  ~ &  ~   &  ~   \\  
5746 & 1677 & 0.16 & 11.00  & 15.57 & 17.43  & 20.55 &  2.144  & 45  &  4.10  &  3 & 0.83 & 0.324\\  
  ~  &  ~   &  ~   &   ~    &   ~   &   ~    &   ~   & (0.025) &  ~  & (0.43) &  ~ &  ~   &  ~   \\  
5838 & 1337 & 0.13 &  6.66  & 14.56 &  8.05  & 18.83 &  2.407  & 86  &  4.04  & -3 & 0.63 & 0.319\\  
  ~  &  ~   &  ~   &   ~    &   ~   &   ~    &   ~   & (0.010) &  ~  & (0.34) &  ~ &  ~   &  ~   \\  
5854 & 1708 & 0.17 &  5.04  & 15.53 &  5.09  & 19.14 &  1.986  & 65  &  4.12  & -1 & 0.70 & 0.191\\  
  ~  &  ~   &  ~   &   ~    &   ~   &   ~    &   ~   & (0.027) &  ~  & (0.47) &  ~ &  ~   &  ~   \\  
5879 & 1065 & 0.10 &  1.90  & 15.68 &  2.99  & 20.11 &  1.761  & 42  &  2.21  &  4 & 0.70 & 0.190\\  
  ~  &  ~   &  ~   &   ~    &   ~   &   ~    &   ~   & (0.060) &  ~  & (0.31) &  ~ &  ~   &  ~   \\  
5965 & 3603 & 0.35 &  6.26  & 15.77 &  8.50  & 20.12 &  2.210  & 46  &  3.29  &  3 & 0.83 & 0.230\\  
  ~  &  ~   &  ~   &   ~    &   ~   &   ~    &   ~   & (0.021) &  ~  & (0.26) &  ~ &  ~   &  ~   \\  
6010 & 1923 & 0.19 &  2.10  & 13.76 &  2.71  & 17.85 &  2.157  & 68  &  2.21  &  0 & 0.77 & 0.297\\  
  ~  &  ~   &  ~   &   ~    &   ~   &   ~    &   ~   & (0.018) &  ~  & (0.34) &  ~ &  ~   &  ~   \\  
6504 & 4680 & 0.46 &  3.86  & 15.21 &  4.21  & 19.35 &  2.266  & 53  &  2.59  &  2 & 0.80 & 0.347\\
  ~  &  ~   &  ~   &   ~    &   ~   &   ~    &   ~   & (0.014) &  ~  & (0.36) &  ~ &  ~   &  ~   \\
7332 & 1550 & 0.15 &  2.99  & 12.97 &  3.66  & 16.91 &  2.051  & 102 &  4.48  & -2 & 0.75 & 0.259\\
  ~  &  ~   &  ~   &   ~    &   ~   &   ~    &   ~   & (0.015) &  ~  & (0.27) &  ~ &  ~   &  ~   \\
7457 & 1114 & 0.11 &  6.68  & 16.50 &  8.28  & 20.53 &  1.749  & 52  &  6.24  & -3 & 0.48 & 0.191\\
  ~  &  ~   &  ~   &   ~    &   ~   &   ~    &   ~   & (0.046) &  ~  & (1.44) &  ~ &  ~   &  ~   \\
7537 & 2717 & 0.26 &  1.28  & 15.69 &  1.52  & 19.62 &  1.619  & 23  &  0.89  &  4 & 0.66 & 0.155\\
  ~  &  ~   &  ~   &   ~    &   ~   &   ~    &   ~   & (0.094) &  ~  & (0.19) &  ~ &  ~   &  ~   \\
\hline
\hline
\end{tabular}}
\end{center}
{\bf \sc{Notes:}}
Photometric errors in columns (5) and (7) are 0.05 and 0.08 mag respectively 
(PB97). For definition of log($\sigma_0$) and Mg$_{2,0}$ see text.
\end{table*}

\subsection{Spectroscopic Parameters}
\subsubsection{Spectral Data Reduction}
\label{Sec:Spectral}
The data were reduced following standard steps using the IRAF
package. Each individual frame was divided by a flatfield, taken at the same 
position, after a bias frame had been subtracted. We subtracted the sky using
the outer regions of the 4$'$ long slit, and removed the cosmic rays 
using the $RED^{UC}_{M}E$ package \cite{c99}. The data were calibrated
in wavelength, with an RMS error in the calibration solution of 0.1 
{\AA} in the blue and 0.02 {\AA} in the red arm. Finally the data 
were calibrated in flux using standard stars BD+284211 \cite{Oke1} and BD+174708
\cite{Oke2}.

\subsubsection{Derivation of kinematic parameters}
\label{Sec:Kinematics}

Kinematic parameters were derived from data in the region of
the IR Ca II triplet (8498 \AA, 8542 \AA, 8662 \AA) in the red 
arm spectra. We worked with coadded spectra covering the central 3 pixels of
each long-slit galaxy spectrum, roughly 1.1 arcsec, which is a good match to the
seeing conditions during the observations. Signal-to-noise ratios per pixel
for these
spectra are listed in Table~\ref{Tab:DataSet}. We made use of the program
FOURFIT developed by van der Marel \& Franx \shortcite{vdm93}, kindly
made available to us by the authors. A Gaussian fit to the line-of-sight
velocity distribution (LOSVD) was performed to obtain mean radial velocities
and velocity dispersions. Star templates of spectral types G8III, K1IIIb, K3III and K5III,
gave dispersions with rms differences around 5 km/s. We took the median
value of the 4 determinations as the best estimate of the central velocity
dispersions of our galaxies. Velocity dispersions results are given in
column (8) of Table~\ref{Tab:DataSet}. 
Aperture corrections have been applied to the central velocity dispersions. We have converted
a rectangular aperture (1.1 x 1.2 arcsec) to an 'equivalent circular aperture'. Then we have
corrected those values to a standard aperture, in this case defined to be 1.7 arcsec 
at the distance of Coma (see J\o rgensen et al. 1995 for more details)

The velocity dispersion errors in Table~\ref{Tab:DataSet} are uncertainty estimates
derived from simulations using FOURFIT. Simulated galaxy spectra were made by broadening
a K3III star with different LOSVD, using Gaussian as well as non-Gaussian profiles, 
and adding white noise to yield S/N of 25, 33, 50 and 100.
These simulated spectra were analyzed with three star templates of types G8III, G9VI and
K3III. For each S/N we define the dispersion error as the rms of all the velocity dispersions
coming from all templates as well as all LOSVDs. The error estimate thus includes
contributions from S/N, template mismatch and non-Gaussian LOSVD. A complete
report of the simulations that we performed will be given in a forthcoming paper
that presents the full LOSVDs of the galaxies.

\subsubsection{Derivation of Mg$_{2,0}$}
\label{Sec:Mg2}
The blue arm data was used exclusively to measure the central Mg$_{2}$ index, here indicated with Mg$_{2,0}$. We have
measured the aperture corrected Mg$_{2,0}$ index (see J\o rgensen et al. \shortcite{j95_2}. 
for more details) and converted it to the Lick system \cite{w94} following the procedure in
Vazdekis \shortcite{v97}.

\subsubsection{Derivation of distances}
\label{Sec:Distances}
Distances to the galaxies were derived from optical recession 
velocities given in RC3 \cite{dv91}, corrected to the centroid of the 
Local Group \cite{kk96}, and assuming a uniform Hubble flow. 
Distance errors are therefore significant, due to 
the Virgocentric flow and the possible influence of the Great 
Atractor. We have taken 150 km/s as the
error in the radial velocity. We have taken this value from the work 
of Aaronson et al. \shortcite{a82} who studied the influence of bulk motions on a sample of spiral 
galaxies at distances comparable to our sample. 

\subsection{Photometric Parameters}
\subsubsection{Derivation of structural parameters in K-Band}
\label{Sec:StrucParsK}
The derivation of the structural parameters in K-band was done as follows.  The 
$K$-band effective radii in arcsec were obtained from APB95. These 
$r_e$ therefore were derived from S\'ersic fits to $K$-band bulge profiles 
after an ellipticity-based bulge-disk decomposition (see APB95 for 
details). The APB95 major-axis $r_e$ values were scaled to the 
geometric mean radius $r_{e}=\sqrt{r_{e,minor} \cdot r_{e,major}}$ 
for direct comparison with other authors (B92, J96, P98) using 
ellipticity profiles from the $K$-band ellipse fits published in 
PB97.

\subsubsection{Derivation of structural parameters in B-Band}
\label{Sec:StrucParsB}
For the $B$-band FP, our goal is to minimize the effects of dust extinction on
the photometry. An $r_{e,B}$ derived from a direct ellipse fit to the $B$-band
image is likely to be affected by the copious dust present on one side of the
galaxy, and also in the nuclear area \cite{p99}. We opt therefore to
derive the dust-corrected $B$-band effective radius $r_{e,B}$ from 
the effective radius in $K$ and the colour gradient determined on the
minor axis. The conversion was done in the following way:
\\
We assume that our profile is of the form:
\\
\begin{equation}
I(r)=I_{e} \cdot 10^{(-b_{n}\cdot([r/r_{e}]^{1/n}-1))}
\label{Eqn:Sersic}
\end{equation}
with $b_{n}=0.868\cdot n-0.142$ \cite{c93}.
Differentiating $B-K=-2.5 \cdot log[I_B(r)/I_K(r)]+cst$
with respect to log(r) gives:

\begin{center}
\begin{equation}
\frac{{\rm d}(B-K)}{{\rm d} ~\log(r)}=\frac{r}{\log(e)} \cdot \frac{{\rm d}(B-K)}{{\rm d}r}
\label{Eqn:Grad}
\end{equation}
\end{center}

Working this expression out, we get the relation between the 
effective radii in the two bands as:

\begin{equation}
r_{e,B}=\left\{1-\frac{\nabla(B-K)}{\delta(n)}\right\}^n\cdot r_{e,K}
\label{Eqn:ReffB}
\end{equation}

where $\nabla(B-K)={\rm d}(B-K)/{\rm d}\log(r)$ is the colour gradient 
taken from PB97 and $\delta (n)=(2.5 \cdot b_{n})/(n \cdot log(e))$. The ratio
$r_{e,B}/r_{e,K}$ does not depend much on the power law used and 
depends mainly on the value of the colour gradient.  Values of $r_{e,B}$ used in 
this paper are derived from eqn.~\ref{Eqn:ReffB}.  

Once the major and minor axis effective radii in both bands are 
known, we can determine the value of $I_{e,B}$ from the $K$-band surface
brightness at r$_{e,major,B}$ and the $B-K$ colour at $r_{e,minor,B}$ 
on the dust-free minor axis. This is
valid  since $B-K$ colour gradients on
the minor axis are small (PB97). To convert from $I_{e,B}$ to the mean surface
brightness within an effective radius we have used the relation:
\begin{equation}
<I_{e}>=10^{[0.030 \cdot (\log[n])^2 + 0.441 \cdot \log(n) + 1.079]} \cdot \frac{I_{e}}{2\pi}
\label{Eqn:Inten}
\end{equation}

\noindent for an r$^{1/n}$ law \cite{c93}. 

\subsection{Presentation of the Data}
\label{Sec:Presentation}

In Table~\ref{Tab:DataSet} we present the data used in the analysis of the
FP. In column (1) we give the New General Catalogue number
\cite{d88}. Column (2) lists the recession velocity of each
galaxy in km/s, corrected to the Local Group \cite{kk96}, from optical
heliocentric velocities given in the RC3. Column (3)
corresponds to the scale in kpc/arcsec. Column (4) gives the $K$-band 
geometric effective radius $r_{e}=\sqrt{r_{e,minor} \cdot r_{e,major}}$ in arcsec. 
Column (5) gives the $K$-band central surface brightness in mag/arcsec$^{2}$ within $r_{e}$, 
also from APB95, the photometric errors were assumed to be 0.05 (PB97). Column (6) 
gives the $B$-band effective radius, in arcsec, from eqn.~\ref{Eqn:ReffB}. 
Column (7) is the $B$-band mean central surface brightness within $r_{e}$ in mag/arcsec$^{2}$, the 
photometric errors were assumed to be 0.08 (PB97). Column (8) is the
logarithm of the aperture corrected central velocity dispersion in km/s and its error
(see Section \ref{Sec:Kinematics}). Column (9) is the S/N per pixel of the spectrum used to
determine the central velocity dispersion. Column (10) shows the parameter $n$ of 
the S\'ersic profile and its error from APB95. Column (11) is the morphological type 
from the RC3. Column (12) is the disk ellipticity of the  galaxy as given by APB95.
Finally we tabulate in column (13) the aperture corrected central Mg$_{2,0}$ index in the Lick 
System \cite{w94}, in magnitudes.

\section{Bulges on the Fundamental Plane}
\label{Sec:FundPlane}
We determine the FP by means of an orthogonal fit to the function,

\begin{equation}
\log(r_{\rm e}) = \alpha \cdot \log(\sigma_{0}) + \beta \cdot <\mu>_{e} + \gamma
\label{Eqn:FP}
\end{equation}

We minimise the sum of the absolute deviations of the points 
orthogonally to the relation, using the program GAUSSFIT \cite{j87}. We have
chosen this method since it is relatively insensitive to the presence of outliers
(J96, P98) and is more robust than minimizing the sum of the squares of the absolute
deviations of the points.

Because of the reduced size of our bulges sample, we are not
obtaining a FP solution for our objects alone.  Rather, we perform fits
to various combinations of our sample and other samples in the
literature, namely those of J96 in the $B$-band, and that of P98 in the
K band. In order to avoid errors due to distance uncertainties 
between clusters, we take only a subsample
of the galaxies of J96 and P98: the Es and S0s in the Coma cluster. 
FP differences when including our objects provide 
then a measure of the deviation of our bulges with respect to the FP 
of ellipticals and lenticulars. We minimise any dependency on the 
fitting method by performing our own fits to the data published by 
B92, J96 and P98, rather than using their FP solutions.

The results of the fits for $B$ and $K$ band data are given in
Table~\ref{Tab:FPfits}. We list each parameter, its error, the number
of objects employed, the rms of the solution and the rms of our bulges in each
fit. We have also included in the last column the dispersion expected from 
observational errors; 
the latter is computed as described in Section 3.1.  Uncertainties in the 
coefficients have been calculated by 
performing a bootstrap analysis with 1000 iterations. 

\begin{table}
\begin{center}
\caption{The Fundamental Plane coefficients}
\label{Tab:FPfits}
{\tabcolsep=2.5pt 
\begin{tabular}{cccccccc}
\hline
\hline
Sample & Band & $\alpha$ & $\beta$ & $\gamma$ & N & rms & ~ \\
~ &  ~ & ~ & ~ & ~ & ~ & rms$_{\rm{Bulges}}$ & $\sigma_{obs}$ \\
\hline
B92      & $B$ &  1.63  &  0.27  & -8.84 & 103 & 0.185 & ~ \\
~        &  ~  & (0.09) & (0.02) &   ~   &  ~  & 0.168 & ~ \\ 
J96(Coma)& $B$ &  1.20  &  0.34  & -9.08 &  28 & 0.061 & ~ \\
~        &  ~  & (0.12) & (0.02) &   ~   &  ~  & 0.150 & 0.061 \\ 
Common   & $B$ &  1.15  &  0.31  & -8.37 &  46 & 0.098 & ~ \\
fit      &  ~  & (0.15) & (0.02) &   ~   &  ~  & 0.136 & 0.060 \\
\hline
P98(Coma)& $K$ &  1.38  &  0.32  & -7.85 & 64 & 0.083 & ~ \\
~        &  ~  & (0.20) & (0.03) &   ~   &  ~ & 0.128 & 0.060 \\
Common   & $K$ &  1.30  &  0.30  & -7.31 & 82 & 0.092 & ~ \\
fit      &  ~  & (0.15) & (0.01) &   ~   &  ~ & 0.125 & 0.063 \\
\hline
\hline
\end{tabular}}
\end{center}
\end{table}

For the $B$ band, the fit corresponding to the B92 sample
excludes the dwarf spheroidals.  ``Common Fit'' denotes the fit to the 28
galaxies of the Coma cluster from J96  [hereafter
J96(Coma)] and our sample. Our 
fitted parameters agree well with the ones given by J96 , with slightly higher 
uncertainties due to to the smaller number of points used.
   
The Coma cluster data from P98  [hereafter 
P98(Coma)] have been used to study the  FP in the $K$-band. We performed
a new FP fit to all 64 galaxies in P98,
instead of the 60 used by the authors, given the lack of information about
which  ones had been excluded by P98.  Our P98(Coma) FP is identical to that
given by  P98, including the errors in the fit.  The ``Common fits'' to the  
FP have the added value of extending the FP of galactic spheroids to low 
log(r$_e$) objects.

\subsection{Error Determination}
\label{Sec:Intrms}

The FP fits are shown in figures \ref{Fig:FPblue} and \ref{Fig:FPred}.  As
usual, the errors in log($r_e$) and $<\mu>_e$ are highly anticorrelated, with a
linear correlation coefficient \rm{r} of about 0.97  \cite{j95_1}, which
implies that the error in the combination that enters the
Fundamental  Plane $ log(r_e)-\beta <\mu>_e$ are much smaller than the
individual errors in log($r_e$)  and $<\mu>_e$. For this reason those errors
are not plotted. The error in the combination $log(r_e) - \beta <\mu>_e$ was
derived from a comparison with other authors (A. Graham, private
communication). This  gives us a mean value of
$\sigma_{\rm{log}(r_{e})-\beta<\mu>_{e}}$=0.036.  The same value has been 
applied for
both bands. The observational error on the FP, including the uncertainties in 
$\log(\sigma)$ and distance is then 
\begin{equation}
\sigma_{obs}^2 = \sigma_{(\log(r_e) - \beta<mu>_e)}^2 + (\alpha \sigma_{\log(\sigma_0)})^2 + \sigma_D^2)
\label{Eqn:Obsrms}
\end{equation} 
and the intrinsic scatter can be determined as 
\begin{equation}
\sigma_{int} = [\sigma_{fit}^2 - \sigma_{obs}^2]^{1/2}.
\label{Eqn:Intrms}
\end{equation}
Observational errors are listed in column 8 of Table 2.

An independent estimate of the measurement errors comes from comparing our
data with the literature. Unfortunately we have no objects in common with P98(Coma)
and J96(Coma), and only one object in common with B92: NGC~7332. For this object,
while the $\sigma$'s are very similar, we measure log($r_{e}/{\rm kpc}$)=$-$0.26$\pm$0.042
whereas B92 give log($r_{e}/{\rm kpc}$)=0.010. For the surface brightness
$<\mu>_{e,B}$ the values are 16.91 and 18.30 respectively. The difference in 
$log(r_e) - \beta <\mu>_e$ between these measurements, 0.70, is much larger than 
the measurement errors in either our or B92's data. 
Since the effective radius of  the bulge of NGC~7332 is small, one
might think that a difference in seeing might have caused the  difference. 
However, our sample contains several smaller bulges,  which are lying very
close to the FP, indicating that differences in seeing  will not move objects
away from the FP.

We decided not to consider NGC~7332 in any of the common fits since it
seems to be a very peculiar object and its observational errors are
not big enough to explain such deviation from the different FP
relations. It hasn't been included either in the studies of residuals
of the FP with other parameters presented in following sections. This galaxy
will be addressed again using, among other techniques, integral field spectroscopy
in future papers.

\subsection{The $B$-band Fundamental Plane}
\label{Sec:BFP}

\begin{figure*}
\begin{center}{\parbox{15.3cm}{\hbox{
\psfig{figure=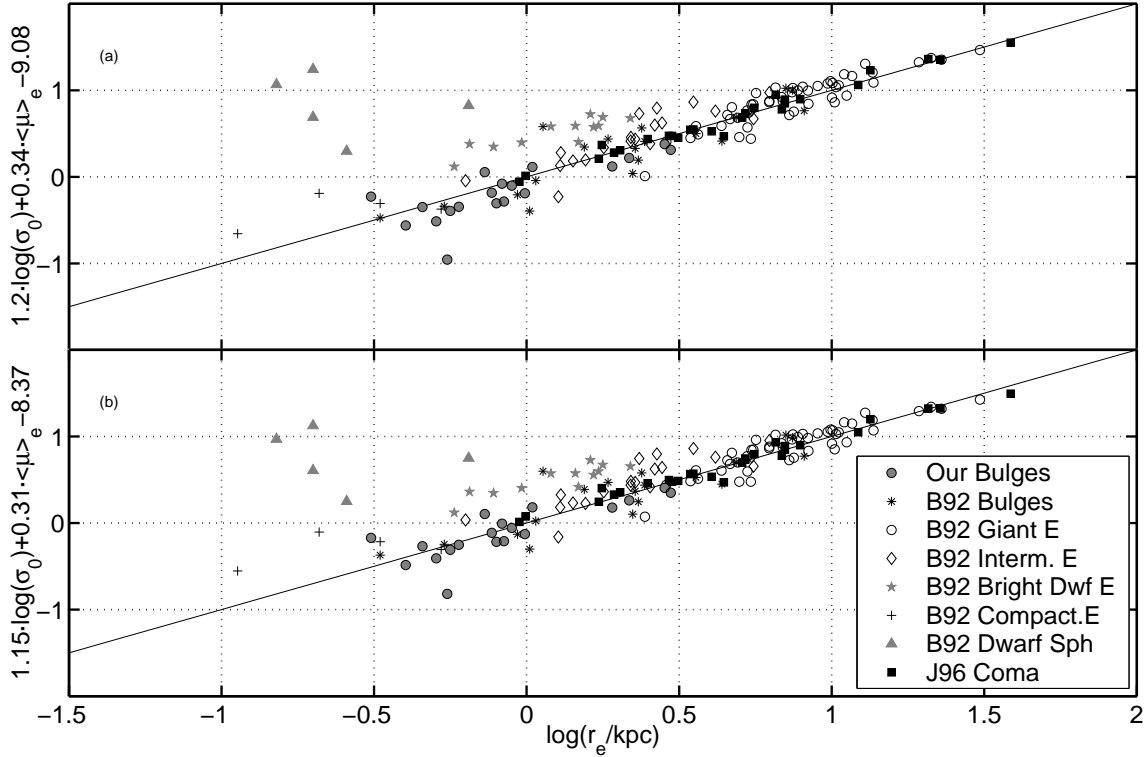,angle=-90,height=4in,width=6in}}}}
\caption{The B-Band FP. Velocity dispersions are in km/s, and surface 
brightness is in B mag/arccsec$^{2}$. We give two different fits to the FP. In
Fig.~\ref{Fig:FPblue}a we have performed a fit to J96(Coma) galaxies and we
have also ploted our bulges and B92 galaxies following J96(Coma) definition. 
Fig.~\ref{Fig:FPblue}b represents a common fit using all the points
(J96(Coma) + our bulges - NGC7332).}
\label{Fig:FPblue}
\end{center}
\end{figure*}

The distribution of galaxies in $B$-band FP space is shown in Fig.~\ref{Fig:FPblue}.  
Two projections are given.  Fig.~\ref{Fig:FPblue}a shows the view orthogonal to the J96(Coma) FP
solution.  Fig.~\ref{Fig:FPblue}b shows the view orthogonal to the
Common Fit FP (J96(Coma) plus this work minus NGC~7332). Substantial
differences are found between the two distributions.  
Figure~\ref{Fig:FPblue}(a) shows
that our bulges fall slightly below the relation defined by the Es and S0s of
J96(Coma). Other classes of objects also show global displacements with respect to the FP,
resulting in the apparent thickening of the galaxy distribution as we
move on to smaller sizes.  To analyse the behaviour of the
different classes of galaxies we have computed the mean separation 
from the FP for each of the object groups, by measuring the mean deviation $\overline x$
along the log($r_{e}$) direction ($x \equiv {\rm object} - {\rm FP} $),
as well as the error in the mean ($\Delta \overline x$).
Deviations from the FP are listed in  Table~\ref{Tab:Devsblue} for each of
the groups in  Fig.~\ref{Fig:FPblue}, and for the two FP solutions (J96(Coma), and 
Common). Aperture corrections are essential to the determination of                
the offset.  Earlier determinations by us of the FP offset of bulges,           
which did not include aperture corrections for $\sigma$, yielded an             
insignificant offset of bulges wrt. J96(Coma) \cite{b02}.

Table~\ref{Tab:Devsblue} shows that our bulges show a 
2-sigma deviation with respect to the J96(Coma) FP. 
The lower $r_{e}$ values for our
bulge  sample require a significant extrapolation of the J96(Coma) FP.   The
offset from the Common fit is insignificant, as expected since the  bulge data
are included in this fit. 
Interestingly, the bulges of B92 show insignificant
deviations with respect to the J96(Coma), in apparent contradiction with B92's conclusion
that bulges preferentially lie below the FP. We can trace the origin of this difference
to the fact that the FP of B92, while very similar to the fits shown in this paper, 
is slightly above the  J96(Coma)
FP. 

The decoupling of the bright dwarf ellipticals from the rest of
galaxies (a 15$\sigma$ result) is remarkable.  This behaviour 
was noticed by B92 who referred to the degree of 
anisotropy as the trigger of the observed offset. The structural
parameters of these objects are probably intermediate between
larger ellipticals and dwarf spheroidals.  The  latter
are known to be structurally different from ellipticals or bulges \cite{m98}.

We have computed the  intrinsic scatter of our sample around the Common
fit, along the log($r_{e}$) direction using equation \ref{Eqn:Intrms}. 
We find $\sigma_{int}$ = 0.122, which is
larger  than what can be accounted for by observational errors (0.060); see
Table 2). This supports the idea of the existence of 
intrinsic scatter, not just for ellipticals, but also for bulges. 
We discuss clues on the origin of the intrinsic scatter in Section 5.1.

\begin{table}
\caption{Deviations from the $B$-band FP}
\label{Tab:Devsblue}
\begin{center}
\begin{tabular}{lccccc}
\hline
\hline
\multicolumn{1}{c}{Group} & \multicolumn{2}{c}{J96(Coma)} & ~ & \multicolumn{2}{c}{Common fit}\\
\cline{2-3} \cline{5-6}     
            & $\overline x$ & $\Delta \overline x$ & ~ & $\overline x$ & $\Delta \overline x$\\ 
\hline
Our Bulges	&  0.074 & 0.033 & ~ & -0.008 & 0.032\\
B92 Bulges	&  0.015 & 0.049 & ~ & -0.026 & 0.045\\
B92 Es		& -0.049 & 0.018 & ~ & -0.060 & 0.018\\
B92 DwfEs	& -0.413 & 0.027 & ~ & -0.406 & 0.026\\
B92 DwfSph	& -1.423 & 0.217 & ~ & -1.340 & 0.206\\
J96(Coma)	& -0.003 & 0.012 & ~ & -0.013 & 0.012\\
\hline
\hline
\end{tabular}
\end{center}
{\bf \sc{Notes:}}
$\overline x$ is the mean deviation in the log($r_{e}$) direction ($x \equiv {\rm object} - {\rm FP} $). $\Delta \overline x$
 is the error in the mean.
\end{table}

\subsection{The $K$-Band Fundamental Plane} 
\label{Sec:KFP}

\begin{figure*}
\begin{center}{\parbox{15.3cm}{\hbox{
\psfig{figure=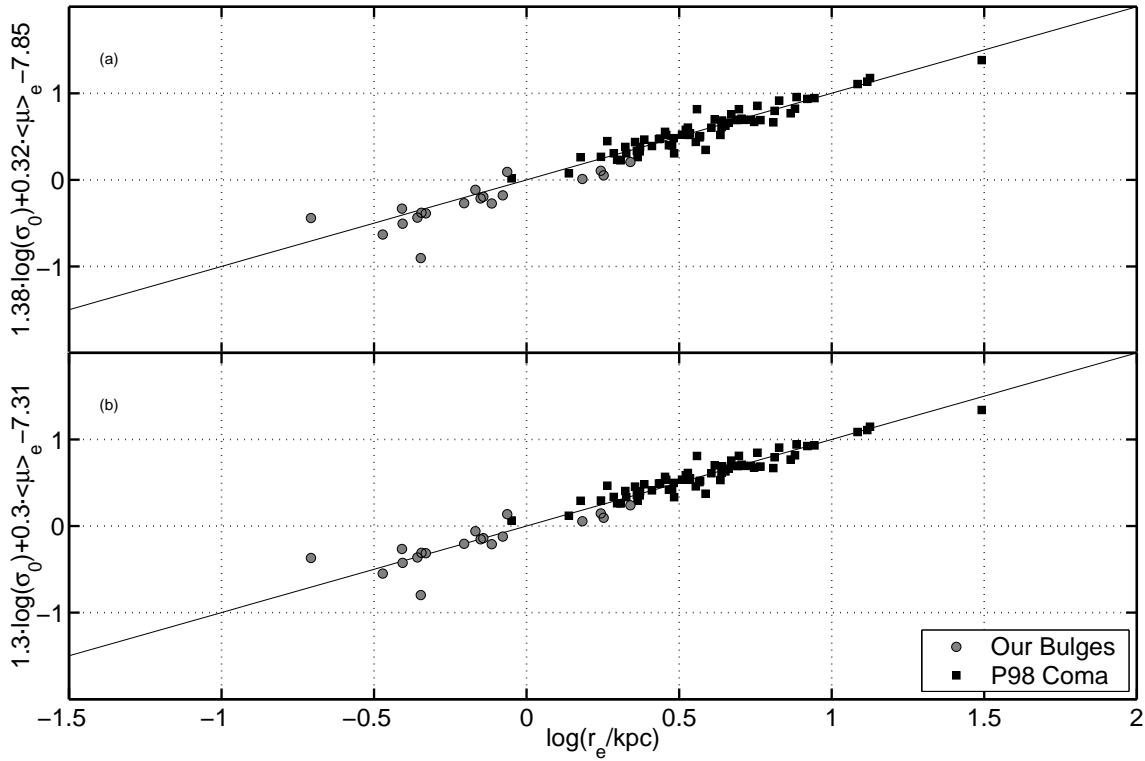,angle=-90,height=4in,width=6in,clip=}}}}
\caption{The K-Band FP. 2 different fits to the FP. In 
Fig.~\ref{Fig:FPred}a we have performed a fit to P98(Coma) galaxies and we
have also plotted our bulges. Fig.~\ref{Fig:FPred}b represents a common
fit using all the points (P98(Coma) + our bulges - NGC7332.).}
\label{Fig:FPred}
\end{center}
\end{figure*}

We have obtained the position of bulges on the $K$-band FP following
the same procedures used for the $B$-band FP.  Our reference elliptical and S0
sample here is the Coma sample of P98 (P98(Coma)). The distribution of galaxies 
in FP space is shown in Fig.~\ref{Fig:FPred}.  As for the $B$-band FP, two 
projections are shown, one orthogonal to the P98(Coma) elliptical FP 
(Fig.~\ref{Fig:FPred}a), and one orthogonal to the Common fit FP (P98(Coma) 
plus our bulges minus NGC~7332; Fig.~\ref{Fig:FPred}b).

Figure~\ref{Fig:FPred} shows that our bulges define a slightly displaced sequence with respect
to the $K$-band FP of Coma ellipticals, extending the FP almost one decade in radius toward 
low-mass spheroids. The most deviant point is NGC~7332.  Its offset with respect to the FP is
unexplained by measurement errors. 

The FP offsets of our bulges with respect to the P98(Coma) and the Common FP's, computed
as described in Section \ref{Sec:BFP}, are given in Table~\ref{Tab:Devsred}. 
The deviations again are significant when using the P98(Coma) FP solution. These
results highlight the displacement of bulges to cluster ellipticals in the $K$ band.

Again, the scatter of our sample around the Common fit in the $K$-Band,
along the log($r_{e}$) direction turns  out to be larger (0.125) than the
dispersion of our bulges on the FP due to uncertainties in the observational
parameters (0.063) (see Section \ref{Sec:Intrms}), implying that we find 
an intrinsic scatter in the $K$-band of 0.108. The fact that our intrinsic
scatter in both bands is almost equal implies that for our sample stellar
population differences are not responsible for most of the scatter, since this
would lead to much larger scatter in $B$.  
We would like to remark that there is a strong paucity of  FP data for
low velocity dispersion galaxies ($\sigma < 100$ km/s), also ellipticals and
S0s, and especially in the
field, a likely consequence of the relatively recent arrival of large-format IR
detectors. Measuring the intrinsic scatter of the FP in the $K$-band 
as well as better constraining any offsets of bulges with respect to the FP of 
ellipticals will require data for significantly larger samples.

\begin{table}
\begin{center}
\caption{Deviations from the $K$-band FP}
\label{Tab:Devsred}
\begin{tabular}{lccccccc}
\hline
\hline
\multicolumn{1}{c}{Group} & \multicolumn{2}{c}{P98(Coma)} & ~ & \multicolumn{2}{c}{Common fit}\\
\cline{2-3} \cline{5-6}     
            & $\overline x$ & $\Delta \overline x$ & ~ & $\overline x$ & $\Delta \overline x$\\ 
\hline
Our Bulges	& 0.054 & 0.029 & ~ & -0.006 & 0.029\\
P98(Coma)	& 0.002 & 0.010 & ~ & -0.007 & 0.010\\
\hline
\hline
\end{tabular}
\end{center}
{\bf \sc{Notes:}}
The mean separation ($\overline x$) has been determined by measuring the mean deviation $\overline x$
along the log($r_{e}$) direction ($x \equiv {\rm object} - {\rm FP} $). $\Delta \overline x$
is the error in the mean.
\end{table}

\begin{figure}
\begin{center}{\parbox{15.3cm}{\hbox{
\psfig{figure=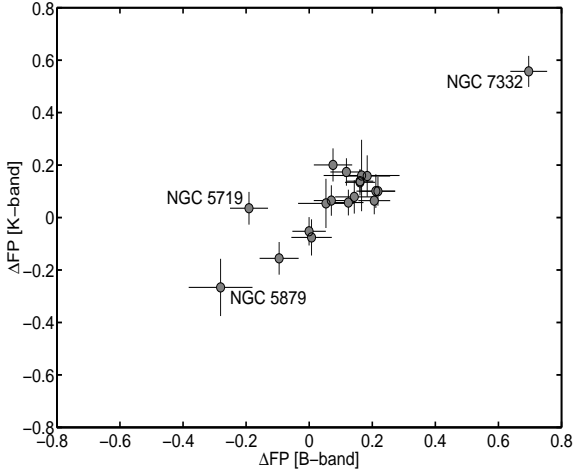,angle=-90,height=2.5in,width=3in,clip=}}}}
\caption{Residuals from the Fundamental Plane in $B$ vs. those in $K$.}
\label{Fig:Residuals}
\end{center}
\end{figure}

In Figure~\ref{Fig:Residuals} we have plotted the residuals of the fundamental 
plane in the $B$-band against the residuals in the $K$-band for the J96(Coma)
and P98(Coma) fits. Residuals are
correlated, something which we expect from the way these quantities were
calculated. The most extreme objects are NGC~7332 and NGC~5879.  A third 
outlier is NGC~5719. If a deviation from the plane is caused by dust,
the object in this diagram will lie on the left of the origin. 
Based on previous work, we don't think that dust extinction is affecting either
of these 2 galaxies. The reason is the way that our fundamental plane parameters
have been determined. Instead of measuring the effective radius and the surface
brightness in $B$ we used the values in $K$ and corrected them using the $B-K$
colour at r$_{e,K}$ and a constant
colour gradient (see Section \ref{Sec:StrucParsB}), determined in the region where 
we decided, based on the flatness of the colour profile, that extinction was not 
important. In PB97 this region was determined to be the region outside 3$''$ for 
NGC~5719 and outside 1.5$''$ in NGC~5879. This method should in principle work as
long as r$_{e,K}$ would lie outside the dusty region. In fact, r$_{e,K}$ = 2.77$''$ 
for NGC~5719 and 1.90$''$ for NGC~5879, both at the border of the dustfree regions,
showing that the FP parameters for these 2 galaxies are barely
affected. If the deviation is due to young stellar populations, the object 
would lie right of the origin. The fact that there are no objects situated here 
shows that the sample does not contain object within significant amounts of 
young stars in the bulge. The fact that for NGC~7332 the deviations are so 
similar shows that the deviations might be related to the details of the stellar
distribution, non-homology, or other factors. For NGC~5879, a small galaxy, the 
error in its velocity dispersion is so large that little can be concluded.
(Section \ref{Sec:Kinematics}).

\section{The M\lowercase{g}$_{2,0}$ - \lowercase{$\log$}($\sigma_{0}$) Relation}

\begin{figure*}
\begin{center}{\parbox{15.3cm}{\hbox{
\psfig{figure=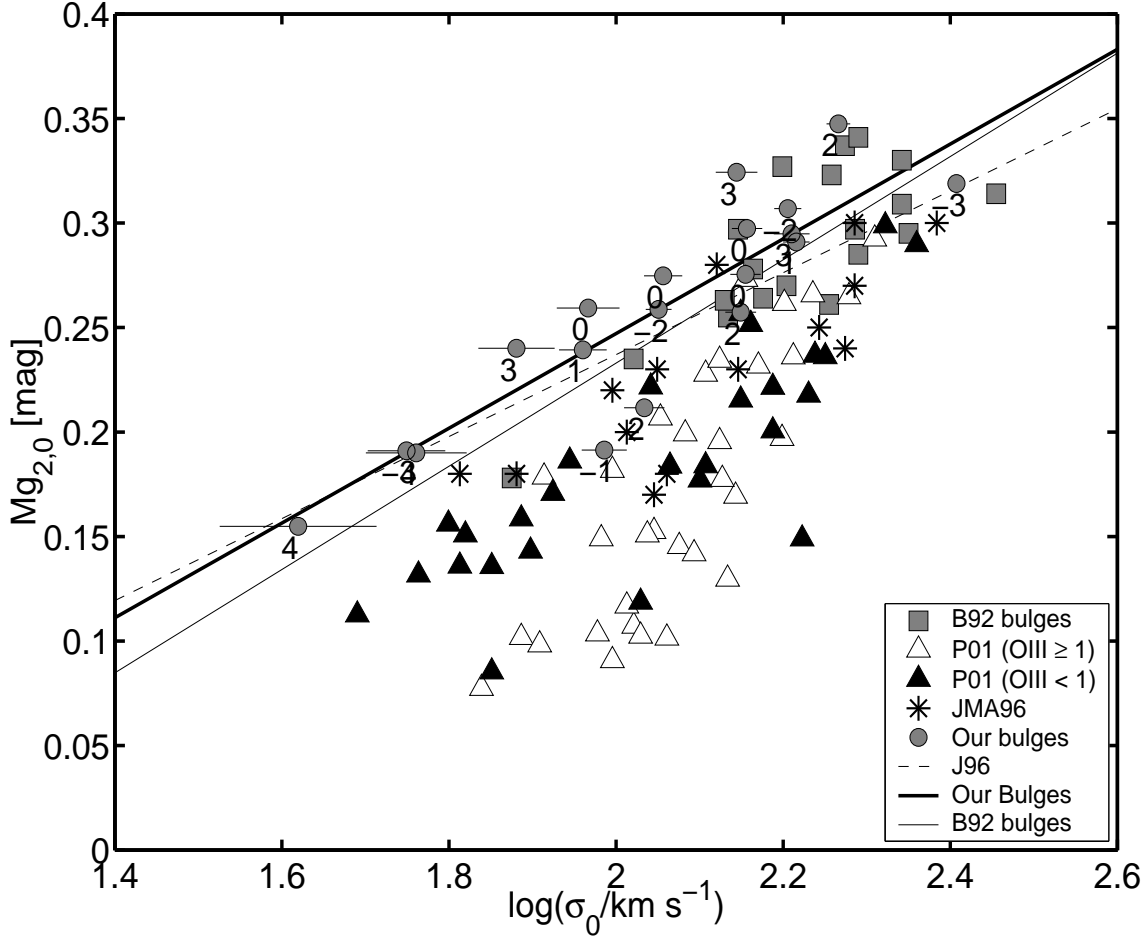,angle=-90,height=5in,width=6in,clip=}}}}
\caption{The Mg$_{2,0}$ - log($\sigma_{0}$) Relation. Filled circles correpond 
to our sample of bulges while solid squares represent the sample of bulges of B92. 
Morphological types of the galaxies of our sample are indicated.
Open triangles are bulges with EW(OIII)$\geq$1, while filled triangles are bulges with 
EW(OIII)$<$1 both from Prugniel, Maubon \& Simien 2001. Asteriscs correspond to
the sample of bulges of Jablonka, Martin \& Arimoto 1996. 
Errorbars of log($\sigma_0$) for our sample are also plotted.}
\label{Fig:Mg2sigma}
\end{center}
\end{figure*}

It is well known that the absorption line index Mg$_{2,0}$
correlates strongly with velocity dispersion \cite{t81,b93}.
This relation can be interpreted as a mass-metallicity dependence  (e.g.
Schweizer \& Seitzer 1992,  Prugniel \& Simien 1996). Since,
according to them, in early-type systems deviations are generally attributed to
the presence of young stars,  this relation can identify young stellar
populations as producers of the scatter around the FP. We have determined this
relation for our sample of bulges as well as for a number of samples in the 
literature (Fig.~\ref{Fig:Mg2sigma}). 

The thick solid line corresponds to a least squares fit to our  points,  while
the thin solid line is a standard least squares fit to the bulges of B92 (solid
squares). The dashed line is the fit given by J96 for Es and S0s. The
coefficients  of the fits are given in Table~\ref{Tab:Mg2sigma}.  We have also
included  the data of Prugniel, Maubon \& Simien (2001, hereafter P01), a sample
of 89 bulges, mostly of spiral i.e. non-S0  galaxies. These bulges can be split
into 2 subsamples: open  triangles are bulges with EW(OIII)$\geq$1, filled
triangles are bulges with  EW(OIII)$<$1. Finally we have also added the
bulges from Jablonka, Martin \& Arimoto (1996, hereafter JMA96) (asteriscs).

We first try to explain the difference between the position of our bulges and the
ellipticals and S0s of J96. In the picture that the Mg$_{2}$ --  $\sigma$
relation describes the dependence of metal enrichment on the depth of the
gravitational potential, the position of bulges above the
relation of  ellipticals may be taken as a horizontal offset to the left --
$\sigma_{0}^{2}$ underestimating the binding energy by not including the
rotation. The contribution of rotation to $\log(\sigma_{0})$ may be estimated
as $0.5\,\log(1 + 0.62  V_{0}^{2}/\sigma_{0}^{2})$ \cite{ps94}. An offset of
0.014 in Mg$_2$ can be brought back onto the J96 relation by an offset of
0.071 in $\log(\sigma_{0})$. A mean V/$\sigma$ of 0.5 suffices to bring
the points back on the Mg$_2$ - $\sigma$  relation. We conclude that non-inclusion of
the rotation alone could explain  the offset of our bulges sample from the
Mg-sigma relation of old ellipticals.

Fig.~\ref{Fig:Mg2sigma} also shows that there are significant 
differences between
the various samples of bulges. 
To establish the nature of them we have compared the observables for
objects in common with these authors (see Table~\ref{Tab:Comparison}). 
For Mg$_2$ the average 
differences are 0.010 $\pm$ 0.006 (RMS) for B93, --0.013 $\pm$ 0.013 for T98
and --0.024 $\pm$ 0.006 for Golev et al. The first 2 are consistent with the
systematic error of the Lick system of 0.008 (Worthey et al. 1994). 
We conclude that the Mg$_2$ values of P01 might be underestimated,  leading to
an offset in Fig.~\ref{Fig:Mg2sigma}.  One should note that, although the data
of Golev et al. (1999) and  Simien \& Prugniel (1997), used in Table 6, were
calculated in the large  Lick-aperture, the data of Fig.~4 have been corrected
to the much smaller physical  aperture of 0.4 kpc, so that the real difference
between our sample and P01 in Fig.~4 is probably larger.  Subtle differences in
sample selection may contribute to the  differences in Mg$_2$ $-$ $\sigma$,
despite the similar Hubble-type distributions and the ranges in $\sigma$ in our
and P01's bulges.   Our sample selection did exclude very dusty bulges
(Balcells \& Peletier 1994, Peletier et al. 1999). This might have biased our
sample toward bulges without ongoing star formation. On the other hand, our
sample is less affected by dust optical depth effects on Mg$_2$ and on $\sigma$, as
well as from disk contamination, which  probably affected the measurements of
P01.   Thus our sample allows more accurate measurements of stellar population
diagnostics, but at the price of potentially excluding the youngest objects. 
We note nevertheless that our two Sbc bulges, at the low-$\sigma$ end of the 
distribution, fall right on the relation of old ellipticals; these objects  are
moderately dusty.   For log($\sigma$) the average differences are  0.066 $\pm$
0.013 (RMS) for B92, 0.084 $\pm$ 0.074 for T98 and 0.070 $\pm$ 0.025 for Golev
et al. 
It appears that our velocity dispersions are consistently lower
that the values in the literature. We think this might be due to the fact that
our dispersions were calculated using minor axis spectra, while all others used
major axis spectra. This means that in our case we did not suffer from possible
rotation, which would tend to broaden the spectra. The shift of 0.07 
in the $\log(\sigma)$ direction proposed above to bring the bulges onto the 
J96 relation would bring our datapoints toward those of P01, but a much larger shift would be required to explain the whole difference.
The data of Jablonka et al. (1996) fall in between us and P01, and could be 
considered consistent with us (given the uncertainties in the zero point of the
Lick system and the above mentioned shift of 0.07 in the log($\sigma$)
direction).
      
Another important point to see from Fig.~\ref{Fig:Mg2sigma} is that the slope of
all three bulge samples is steeper than that of J96 for Es and S0s. The
traditional way to explain this is that low velocity dispersion galaxies contain
a larger fraction of young stellar populations (P01). This can be confirmed if,
e.g., the galaxies furthest away from the FP of ellipticals have the latest
morphological type. For our sample the answer is not clear. While the 2 galaxies
with type 4 are situated the most below the FP, no trend is visible for the
others. For the bulges of P01, the Mg$_{2,0}$ deficit  with respect to
ellipticals is very pronounced, especially so for  galaxies  with strong [OIII]
emission, indicating significant amounts of ionised gas.  If one assumes that 
ionised gas traces ongoing star formation one can 
understand their position in the Mg$_2$ $-$ $\sigma$ relation as due to young
stars diluting the Mg$_2$ index. For this reason P01 called the Mg$_{2,0}$ --
$\sigma$ relation a mass-metallicity  relation for bright ellipticals, becoming
gradually a mass-age relation as one goes to fainter objects. 
Under this view of the  
Mg$_2$ $-$ $\sigma$ relation, Figure~\ref{Fig:Mg2sigma} 
is interpreted as indicating that 
the young-star contents of our bulges is much lower than in the P01 sample,
and comparable to that of S0's of B92.  If indeed low Mg$_2$ values are due 
to dilution by a young stellar continuum, the Mg$_2$ $-$ $\sigma$ 
distribution for our bulges argues against a mass-age relation for small 
objects.

\begin{table}
\begin{center}
\caption{The Mg$_{2,0}$ - log($\sigma_0$) Relation}
\label{Tab:Mg2sigma}
\begin{tabular}{lcccc}
\hline
\hline
\multicolumn{1}{c}{Group}& a & $\Delta$a & b & $\Delta$b\\
\hline
J96       	 & 0.196 & 0.016 & 0.257 &   ~  \\
B92 Bulges	 & 0.247 & 0.045 & 0.258 & 0.008\\
Our Bulges	 & 0.237 & 0.030 & 0.271 & 0.006\\
P01 (OIII$\geq$1)& 0.447 & 0.060 & 0.183 & 0.007\\
P01 (OIII$<$1)	 & 0.238 & 0.033 & 0.201 & 0.006\\
\hline
\hline
\end{tabular}
\end{center}
{\bf \sc{Notes:}}
Different fitting coefficients of the Mg$_{2,0}$ - log($\sigma_{0}$) relation,
defined as Mg$_{2,0}$=a $\cdot$ [$\log$($\sigma_{0}$) $-$ 2.1]+b
\end{table}

\begin{table}
\begin{center}
\caption{Comparison with other authors}
\label{Tab:Comparison}
{\tabcolsep=4.pt
\begin{tabular}{lccccc}
\hline
\hline
\multicolumn{1}{c}{Galaxy} & Parameter & Us & B93 & T98 & G99/SP97\\
\hline
NGC~5422 & Mg$_2$ & 0.307 & 0.323 &   ~   &   ~   \\
NGC~5854 & Mg$_2$ & 0.191 &   ~   & 0.160 & 0.174 \\
NGC~7332 & Mg$_2$ & 0.259 & 0.263 & 0.253 & 0.225 \\
NGC~7457 & Mg$_2$ & 0.191 &   ~   & 0.189 & 0.171 \\
\hline
NGC~5422 & log($\sigma_0$) & 2.205 & 2.258 &   ~   &   ~  \\
NGC~5854 & log($\sigma_0$) & 1.986 &   ~   & 2.170 & 2.017\\
NGC~7332 & log($\sigma_0$) & 2.051 & 2.130 & 2.114 & 2.117\\
NGC~7457 & log($\sigma_0$) & 1.749 &   ~   & 1.756 & 1.863\\
\hline
\end{tabular}}
\end{center}
{\bf \sc{Notes:}}
B93: Bender, Burstein, \& Faber 1993.\\
T98: Trager et al. 1998.\\
G99: Golev et al. 1999 [for Mg$_2$]\\
SP97: Simien \& Prugniel 1997 [for log($\sigma_0$)]
\end{table}

\section{Discussion}

In the past bulges of spirals have been difficult galactic
components to study. This has been due to the presence of dust, the accompanying
disk, and the fact they are generally small, so that very often ground-based
observations are affected by lack of resolution.
The appearance of the Hubble Space Telescope in the 1990s has vastly 
improved the situation. 
Peletier et al. \shortcite{p99} performed a detailed study of galactic bulges using
NICMOS observations, giving as a result a much more detailed insight of ages
and dust in these objects than was known before. The main conclusions from this
work are that early-type bulges (S0-Sb) behave like ellipticals and seem to be 
as old as cluster Es with very little spread in age ($\sim$ 2 Gyr). This supports the
hypothesis of an early formation for bulges and ellipticals, via either dissipational
collapse or early mergers. Late-type bulges (Sbc onwards) show hints of youth
with respect to the early-types and could belong to a class of objects formed or
transformed by secular disk evolution. In general they are smaller,
younger and have lower central surface brightness. The question then is whether
they fall on the same Fundamental Plane decribed by Es and S0s.

We have measured the mean separation of our bulges with respect to the FP of Es
and S0s (as defined by the Es/S0s of J96(Coma)) and found a small but
non-negligible offset of 0.074 $\pm$ 0.033 in the $B$-band 
and 0.054 $\pm$ 0.029 in the  $K$-band. 
In agreement with other authors (J96,P98) we find that the
FP has intrinsic scatter, not explained by observational
uncertainties. 

The strong proximity of bulges to the FP of ellipticals and S0s reinforces the conclusions of
Peletier et al. \shortcite{p99} that early-to-intermediate bulges are homogeneously old and
structurally similar to elliptical galaxies. It is useful in any event to investigate the origin
of the intrinsic scatter and the offset.  We first analyse the intrinsic scatter by showing the
FP residuals with respect to various galaxy observables.  Then, we discuss the roles of
populations, kinematics and homology in creating the offset with respect to Coma ellipticals.

\subsection{Analysis of the Residuals}
\label{Sec:Residuals}

The existence of this intrinsic scatter has been a matter of
debate for the last decade. For ellipticals, 
Mobasher et al. \shortcite{mob99} have proposed the
population of AGB stars as one of the sources of the intrinsic scatter of the FP
in the K-band. Structural and dynamical non-homology may be responsible, at
least in  part, for the intrinsic scatter \cite{g97}. Recent studies suggest
that  the intrinsic scatter is probably a combination of age/metallicity 
variations and dynamical deviations from homology (P98 and references therein).
For bulges differences in stellar populations might account for even larger
scatter (e.g. Prugniel et al. 2001).

To obtain clues on the origin of the intrinsic scatter of our bulges in 
the FP, we analyse correlations between the residuals of the FP, defined 
as $\Delta FP \equiv \log(r_{e})$-fit, using the
Common fits, and different observables for our sample of bulges:
$\Delta Mg_{2}$, Galaxy Type, inclination of the galaxy and S\'ersic index $n$. 
These correlations should give us ideas whether the distribution of
our bulges along the sequence described by Es and S0s depends on the presence
of young stars [$\Delta FP - \Delta Mg_2$], homology or light
profile shape [$\Delta FP$ - Galaxy Type, $\Delta FP$ - Index of the
S\'ersic profile (n)] or the degree of anisotropy of the object
[$\Delta FP$ - Inclination of the galaxy (i)].  

\begin{figure*}
\begin{center}{\parbox{15.3cm}{\hbox{
\psfig{figure=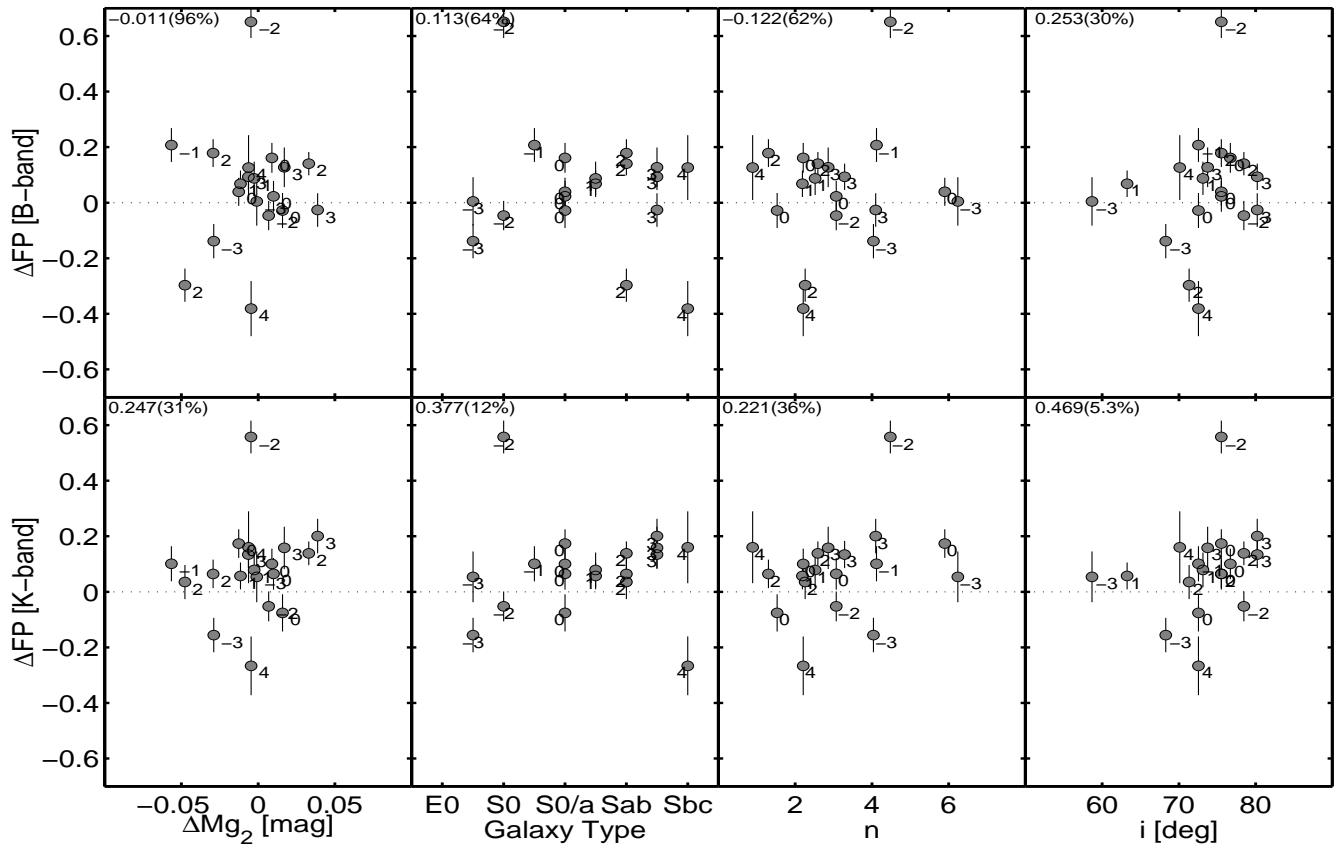,angle=90,height=4.5in,width=7in,clip=}}}}
\caption{Correlations of the $\Delta FP=log(r_{e})-fit$ with global 
galaxy parameters. Points are labeled with the morphological type of the galaxy T.  
The number at the top left of each panel is the Spearman rank correlation
coefficient. The number within parenthesis is the probability of a null
correlation.}
\label{Fig:ResidGlobPars}
\end{center}
\end{figure*}

We perform this study for the 2 bands using the Spearman rank order
correlation coefficients in each case. We find that there is no correlation between
the residuals of the FP and any of the 3 FP parameters in either B or K, 
as can be expected from Section \ref{Sec:FundPlane}.
Figure~\ref{Fig:ResidGlobPars} shows the 
distributions of FP residuals as a function of 
global galaxy parameters in the 
$B$ and $K$-band. Points are labeled with the morphological type of the galaxy 
T. The probability of a null correlation is given within parenthesis in each 
panel, next to the correlation coefficient. The point labeled "-2" near the top 
of each panel corresponds to NGC~7332. This point is not included in the computation 
of correlation coefficients (see Section \ref{Sec:FundPlane}).

No signs of correlations are found of $\Delta FP$ and $\Delta
Mg_2$ (96\% and 31\%, in $B$ and $K$ band respectively). The same is 
true if the deviations are taken from the fit of J96. If the deviations,
both to the FP and the Mg$_2$ -- $\sigma$ relation, are due to young stellar 
populations, or even to non-inclusion of rotational support, a correlation 
is expected (P01). However, for our sample the deviations from both relations
are small, and this fact combined with the relatively small size of the sample
makes us not see the correlation. 

Weak trends are found with galaxy type (64\% and 12\%) in both bands. While weak, 
the distribution of galaxy types (within our bulges) gives a suggestion that the 
FP residuals may depend on galaxy type, in the sense that late type bulges tend 
to be below the FP, and early types above. 

We don't find a relation between $\Delta FP$ and $n$, the S\'ersic
index. The same conclusion  was reached by Graham \& Colless \shortcite{g97}
for Virgo cluster galaxies studied  in the V band at high S/N. Such a 
correlation, however, could be hidden, in  part, by stellar populations
effects. Prugniel \& Simien \shortcite{ps97} found a correlation between the
residuals of the FP and $n$ after removing the scatter indiced by the stellar
populations. Here again think the small size of the sample prevents us from
seeing this correlation. 

Next we analyse possible correlations between $\Delta FP$ and the inclination
of the galaxy (i). Projection 
effects and the presence of disks are both likely sources of scatter in the FP \cite{s93}. 
Despite the limited range in ellipticities, 
we find that the probability for a significant correlation is
70\% and 95\%, in the $B$ and $K$ bands respectively. 
J96 found that galaxies with a large ellipticity (presumably
edge-ons) have a slight tendency to lie below the FP. The residual, for 
$\epsilon$ = 0.6 - 0.8 $\approx$ ~0.10 in log($r_e$), similar to ours.  
The $\Delta FP$ dependence on inclination can contribute to the global
offset of our inclined sample w.r.t. the J96(Coma) FP.  
We address this in Section \ref{Sec:Offsets}.



\subsection{Offsets with respect to ellipticals}
\label{Sec:Offsets}

Population age/metallicity differences, internal kinematics and non-homology each can in
principle shift bulges with respect to the FP of old ellipticals.   We discuss each in turn.  

The position of bulges below the FP of ellipticals could have been produced if bulges are on average
younger than elliptical galaxies  \cite{ss92,ps96}. Taking  the models of
Vazdekis et al. (1996) one finds that for solar metallicity a  decrease in age
from 12.5 to 10.0 Gyr gives a shift in the log(r$_e$) direction  of 0.06 in $B$
and 0.04 in $K$, consistent with the results presented here. This  means that
possibly the offset could be explained if our sample on the average  would be
about 2.5 Gyr younger than the Coma cluster samples used by J96 and P98. 

The effect of rotational support, although small, has been detected by other authors \cite{b92,ps94}.
Following Prugniel \& Simien \shortcite{ps94}, the rotational kinetic energy may be introduced in the FP via a 'rotational support term' $S$  to be added to $\log(\sigma_0^2)$.  Using our definition of FP and after accounting for the inclination range for our sample, we approximate $S$ as 

\begin{equation}
\label{Eqn:S-term}
S = \alpha\,0.5\,\log(1 + 0.62\frac{V_{max}^2}{\sigma_0^2})
\end{equation}

\noindent with $\alpha$ is from Eqn.~\ref{Eqn:FP}.
Since our spectra were obtained along the minor axis of the galaxies, we have
not been able to measure $S$ for our sample.  We can nevertheless estimate the
amount of rotation needed to generate the observed offset.  Non-inclusion of
the rotation term offsets points {\sl downward} on the FP diagram.  A sample
average of $V_{max}/\sigma_0 = 0.66$ produces the observed FP offset.  Within
this model, the observed offset of our bulges sample with respect to Coma
ellipticals can be explained by the non-inclusion of rotation data for our
bulges.    

The effect of non-homology, i.e. that bulges do not all follow the $R^{1/4}$
law, has not been found in our data, given that $\Delta FP$ does not correlate
with $n$, the S\'ersic index.  As noted in Section \ref{Sec:Residuals}, Prugniel \&
Simien \shortcite{ps97} did find such a correlation once the scatter by
population effects had been taken into account.  

Finally, in Section \ref{Sec:Residuals} we have identified galaxy inclination as a
potential origin for FP offsets.  J96 model the inclination effect by
considering the effect of rotation on the measurement of the central velocity
dispersion.  This effect is independent of the rotation support term described
above.  

The previous analysis suggests that any combination of age and kinematics contribution to the FP
can explain the offsets of our sample with respect to Coma ellipticals, whereas we fail to
identify the effects of non-homology.  Most plausibly, not a single one of the two processes is
responsible for the entire offset.  The latter probably results from a combination of the two. 
Nevertheless, we note that, while the population contribution rests on a model determination of
ages, the kinematic offset is sure to be there; indeed, bulges rotate and our sample is selected
to be inclined.  

\section{Conclusions}

We have analysed the position of bulges of early-type spirals (type S0-Sbc) on 
the Fundamental Plane of early-type galaxies. A few conclusions can be drawn from 
our study:

\begin{itemize}
\item{We find that our sample of bulges early-type spirals lies slightly  below
the FP, as defined by E and S0 galaxies, both  in the $B$ and $K$ bands. The
fact that bulges lie so close to the FP of  ellipticals and S0s implies that
their formation epoch must have been similar to  that of cluster Es and S0s.}

\item{Both age and the unaccounted rotation effects are likely to contribute to the offset observed. 
Any single one of these effects can account for the observed offset, by assuming either an age on
average 2.5 Gyr younger than the cluster  ellipticals and S0s, or by assuming an average
$V_{max}/\sigma_0\sim 0.66$.  Because rotation must contribute to the FP offset, the age
differential between bulges and cluster ellipticals must be significantly smaller than the
mentioned 2.5 Gyr.}  

\item{We find that there is a hint that bulges of later morphological type are
situated  below the other bulges in our sample, indicating that they have 
slightly younger ages, again consistent with our results from HST colours.}

{\item We confirm the result of Prugniel et al.\ (2001) that bulges 
have a steeper Mg$_{2,0}$ -- log($\sigma$) relation than ellipticals or S0s. 
Contrary to current lore, the Mg$_{2,0}$ -- $\sigma$ 
is not a mass-metallicity relation only, at least for bulges:  
Younger objects seem to increase the slope of the relation (P01).
For our sample, however, the slope is shallower than for P01, much closer
to that of the ellipticals. }

\end{itemize}

The William Herschel Telescope is operated on the island of La Palma by the
Isaac Newton Group in the Spanish Observatorio del Roque de los Muchachos of
the Instituto de Astrof\'\i sica de Canarias.
We acknowledge the professional help of the observatory 
staff in the operation of the WHT and the ISIS spectrograph.
Dr. Inger J\o rgensen is thanked for giving us data in computer readable form. We
are indebted to the anonymous referee for very useful comments that have help us to
improve this paper.

\end{document}